\documentclass[a4paper]{jpconf}
\usepackage{graphicx}
\usepackage{fancyhdr}
\usepackage{enumerate}
\usepackage{amsmath}
\usepackage{color}
\usepackage{setspace}
\usepackage{amssymb}
\usepackage{dcolumn}
\usepackage[squaren]{SIunits}
\begin{document}

\title{HARPO: beam characterization of a TPC for gamma-ray polarimetry and high angular-resolution astronomy in the MeV-GeV range}

 \author{Shaobo Wang}
 \address{LLR, Ecole Polytechnique, CNRS/IN2P3, France}
 \ead{wang@llr.in2p3.fr}

\author{Denis~Bernard, Philippe~Bruel, Mickael~Frotin, Yannick~Geerebaert, Berrie~Giebels, Philippe~Gros, Deirdre~Horan, Marc~Louzir, Patrick~Poilleux, Igor~Semeniouk}

\address{LLR, Ecole Polytechnique, CNRS/IN2P3, France}

\author{David~Atti\'e, Denis~Calvet, Paul~Colas, Alain~Delbart, Patrick~Sizun}

\address{Irfu, CEA-Saclay, France}

\author{Diego~G\"otz}

\address{AIM, CEA/DSM-CNRS-Universit\'e Paris Diderot, France}

\address{Irfu/Service d'Astrophysique, CEA-Saclay, France}

\author{Sho~Amano, Takuya Kotaka, Satoshi~Hashimoto, Yasuhito~Minamiyama, Akinori Takemoto, Masashi~Yamaguchi, Shuji~Miyamoto}

\address{LASTI, University of Hy\^ogo, Japan}

\author{Schin Dat\'e, Haruo Ohkuma}

\address{JASRI/SPring8, Japan}

\begin{abstract}
A time projection chamber (TPC) can be used to measure the
polarization of gamma rays with excellent angular precision and
sensitivity in the $\mega\electronvolt$-GeV energy range through the
conversion of photons to e$^+$e$^-$ pairs.

The Hermetic ARgon POlarimeter (HARPO) prototype was built to
demonstrate this concept. It was recently tested in the polarized
photon beam at the NewSUBARU facility in Japan. 
We present this data-taking run, which demonstrated the excellent
performance of the HARPO TPC.
\end{abstract}

\section{Introduction}

Gamma-ray astronomy is the primary means by which we can study the
non-thermal processes occuring in cosmic sources such as active
galactic nuclei (AGN), pulsars and gamma-ray bursts (GRBs)~\cite{GRB}.
The ability to measure the linear polarization in this energy range
would provide a powerful diagnostic for the understanding of the
physical processes at work in these sources~\cite{Blazars}.
A telescope with the ability to perform polarimetry above
$1\,\mega\electronvolt$ has never been flown in space
\cite{Forot:2008ud}.
Gamma-ray telescopes have improved in sensitivity and resolution from
COS-B~\cite{COSB} to the Fermi-LAT~\cite{Fermi}, but the performance
of these pair-creation telescopes decreases at low energy where the
background is problematic and the angular resolution is limited by
multiple scattering.

\begin{figure}[ht]
  \begin{center}
    \includegraphics[width=0.5\linewidth]{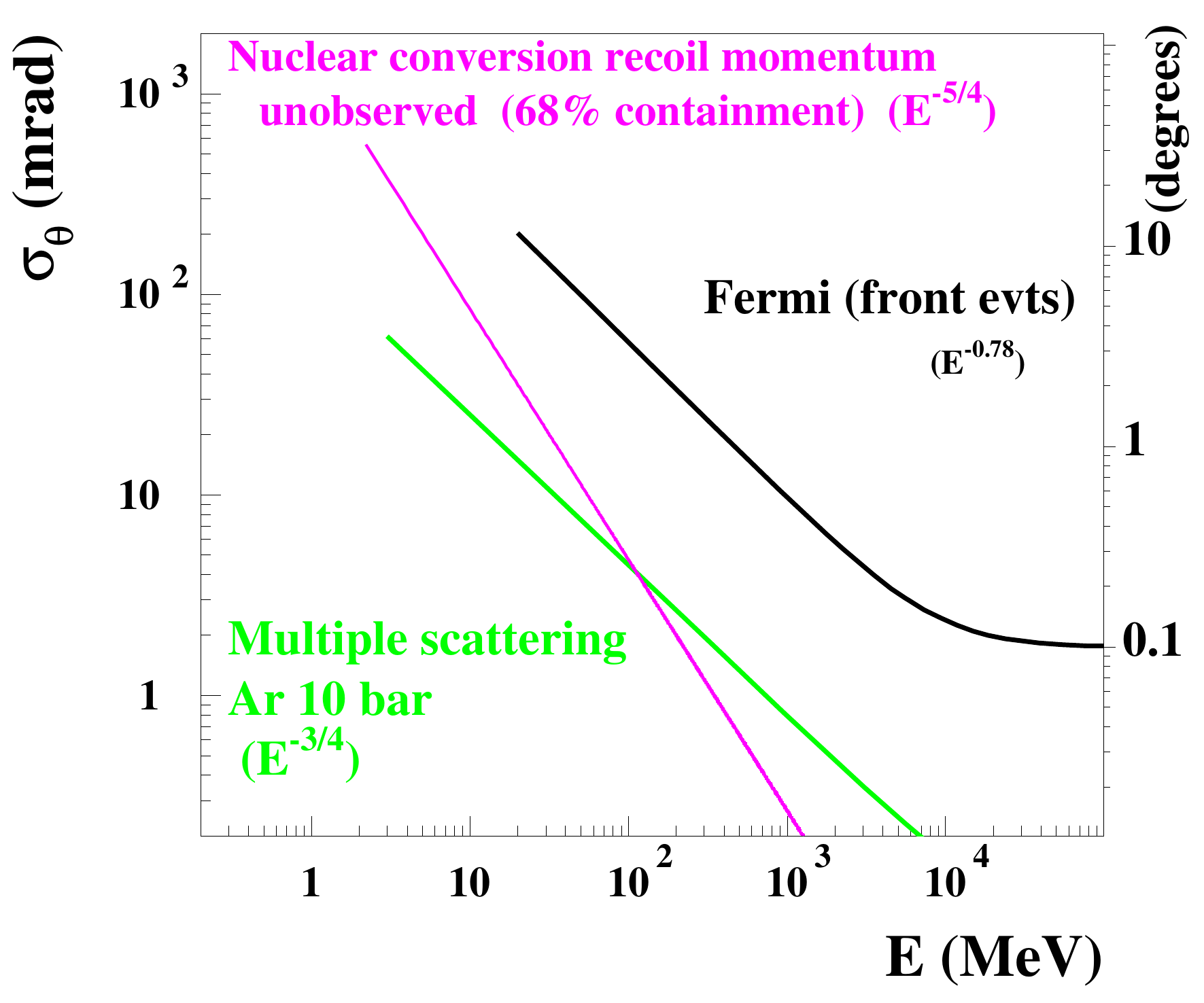}
   \caption{
Various contributions to the angular resolution of a TPC gamma telescope as a function of the photon energy~\cite{Bernard:2012uf}, compared to that of the Fermi-LAT~\cite{Fermi}. The multiple scattering contribution (green) is given for a $10\,\bbar$ argon TPC, using an optimal tracking such as that implemented in a Kalman filter, and with a track sampling of $1\,\milli\meter$ and a spatial resolution of $0.1\,\milli\meter$.
      At high energies ($>100\,\mega\electronvolt$), it is limited by the multiple scattering of the electrons in the gas.
      At lower energies, it is constrained by the unknown recoil momentum of the nucleus.
      This limit can be overcome by looking at triplet conversion, but they are rarer and more difficult to reconstruct.
    }
    \label{fig:angularResAr}
  \end{center}
\end{figure}

The performance and polarimetry potential of a TPC have been studied
in detail in Ref.~\cite{Bernard:2012uf} and~\cite{Bernard:2013jea}.
Several effects contribute to the angular resolution, which is shown as a
function of photon energy in Fig.~\ref{fig:angularResAr} for an argon
gas TPC. The angular resolution for pair production is limited by multiple
scattering of the electrons in the gas. 
Also, most conversions take place in the field of a nucleus whose recoil
momentum is impossible to measure because of the very short recoil
path length.
This puts a limit on the resolution at energies $<100\,\mega\electronvolt$. 
Even with these limitations however, an improvement in
angular momentum of up to an order of magnitude with respect to that
of the Fermi-LAT is achievable with a TPC.

The polarimetry of cosmic sources with a pair-creation telescope has
long been considered to be difficult or even impossible, as thick, high-$Z$
detector elements are needed to convert the incoming photon, and the
conversion electrons then undergo multiple scattering in the
converter, after which the information about the azimuthal angle of the
conversion plane is blurred~\cite{Mattox}.
We have built and validated an event generator~\cite{Bernard:2013jea}
which is full (5D, either nuclear or triplet conversion) exact down to
threshold and polarized.
We have characterized the properties of a telescope based on a thin
homogeneous detector with optimal tracking and have established the
power laws that describe the dominating contributions to the angular
resolution~\cite{Bernard:2012uf}.
For such a detector, even when the dilution of the
effective polarization asymmetry due to multiple scattering is taken
into account, polarimetry can still be performed with high precision
(still under the assumption of optimal
tracking)~\cite{Bernard:2013jea}.

The HARPO project aims to characterize the TPC technology as a high
angular resolution polarimeter and telescope in the MeV-GeV range,
enabling us to obtain unprecedented sensitivity for the detection of low-energy gamma rays.
A TPC is a detector in which traversing charged particles ionize the detector material. The electrons
produced drift along an electric field, $E$,  and are then
amplified and measured on the $x$-$y$ readout plane. The drift time gives
a measure of the $z$ coordinate.
HARPO would be the first space polarimeter above
$1\,\mega\electronvolt$.

 In this paper we first describe the demonstrator that
we have built to validate the performance of the TPC technology, in
particular the characterization of the GEM and micromegas combinations used
for gas amplification
\footnote{
The 2012 tests showed that with the 0.4\,mm narrow collecting
strips that we are using, the micromegas alone does not provide
sufficient amplification at 4 bar for routine operation in a safe
configuration \cite{HARPO:Pisa2012}. 
We have therefore complemented the micromegas with two
layers of Gas Electron Multiplier (GEM).
}.
Then we  present the recent experimental campaign, in which the
detector was exposed to the quasi-monochromatic and almost fully
polarized beam provided by the BL01 line of the NewSUBARU facility,
operated by the LASTI in University of Hy\^ogo in Japan.

\section{The HARPO TPC demonstrator}

\begin{figure}[ht]
  \begin{center}
    \includegraphics[width=0.45\linewidth]{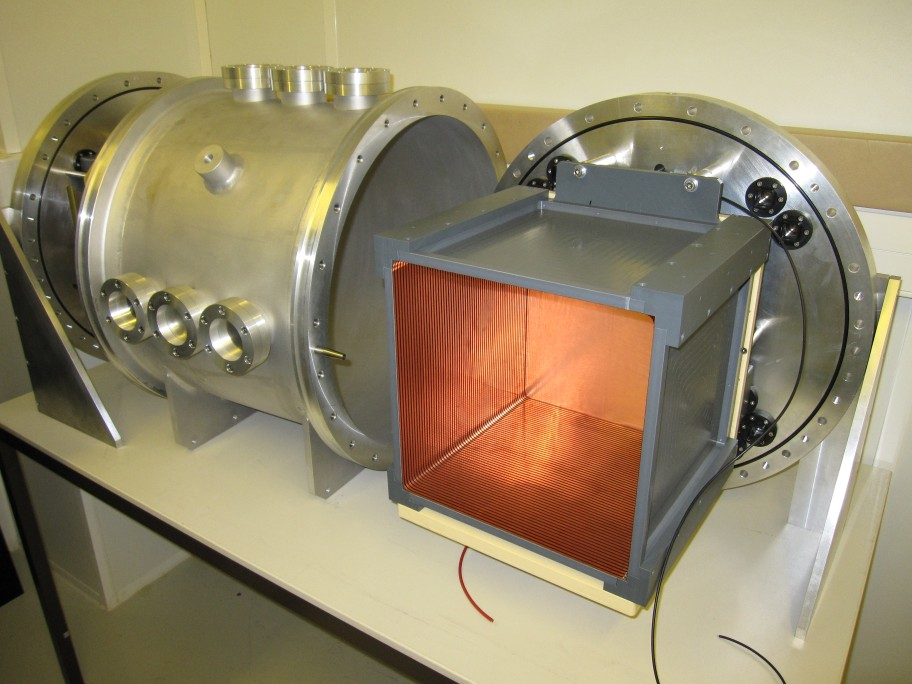}
    \includegraphics[width=0.49\linewidth]{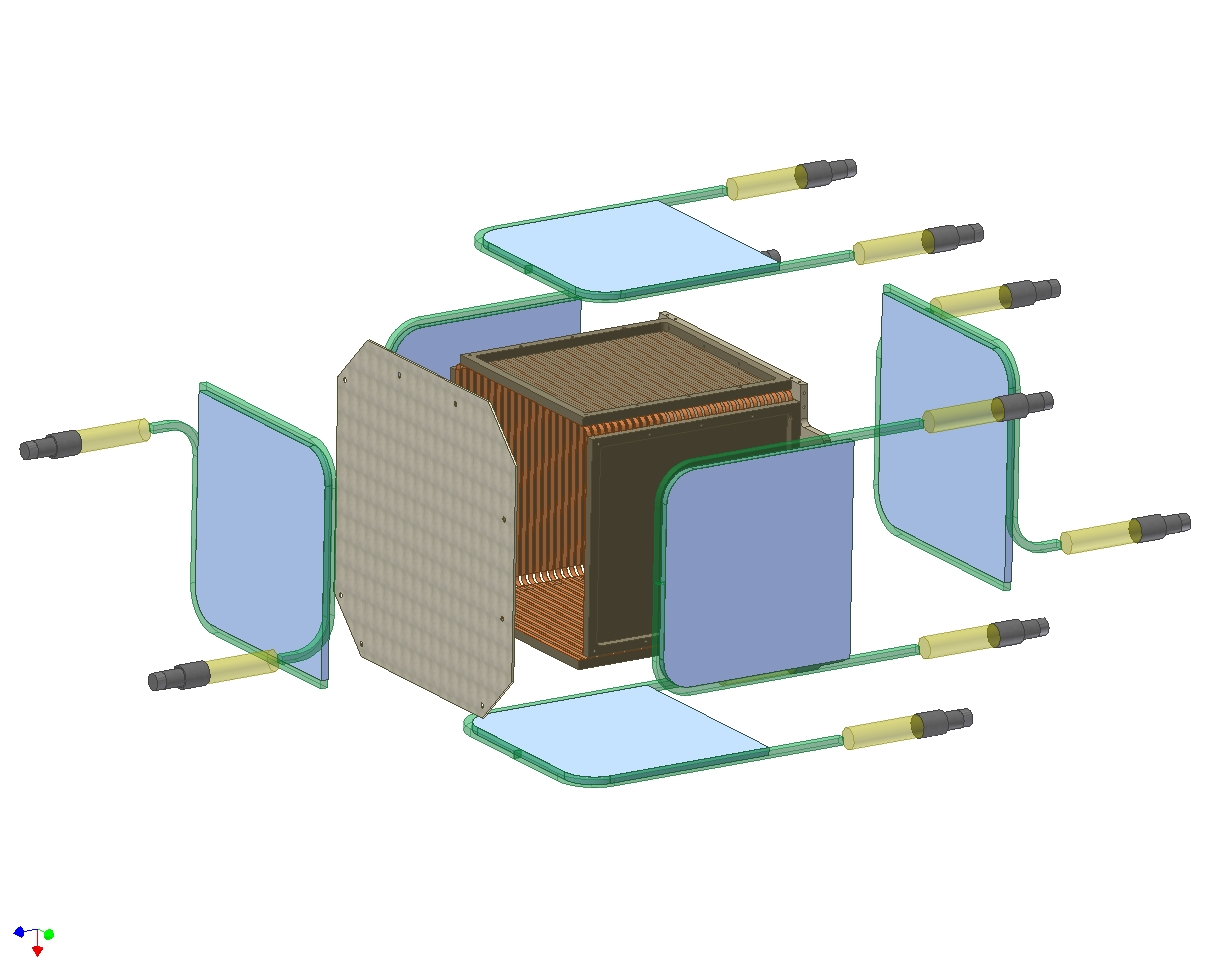}
    \caption{
      Left: the HARPO TPC demonstrator.
      Right: sketch of HARPO. 
      A cubic TPC in the center, with a readout plane of micromegas and GEM on the left.
      Six scintillator plates, each equipped with a wavelength shifter and two photo-multipliers, are used for trigger.
      The system fits in an aluminum cylinder which can operate up to $5\,\bbar$ pressure.
    }
    \label{fig:HARPO}
  \end{center}
\end{figure}

The HARPO demonstrator~\cite{HARPO:Pisa2012} (Fig.~\ref{fig:HARPO}) is a 
$30 \, \centi\meter$ cubic TPC which can be operated from low pressure
up to $5\,\bbar$.
It is surrounded by 6 scintillator plates that provide an external
trigger.
Amplification is performed with the combination of two GEMs~\cite{GEM}
and one micromegas~\cite{micromegas}.

The readout comprises two series ($x, y$, perpendicular to each other) of 288
copper strips with $1\,\milli\meter$ pitch.  
Signals are acquired with AFTER chips~\cite{AFTER} using the Feminos
system~\cite{Feminos}.

\section{The HARPO amplification system}

We first characterized the performance of the GEM+micromegas
amplification system using a radioactive source. The amplification system was subsequently integrated into the TPC detector, where it was
characterized using cosmic rays.

\subsection{Characterization of micromegas and GEM(s) combinations with a radioactive source}

\begin{figure}[ht]
  \begin{center}
    \includegraphics[width=0.33\linewidth]{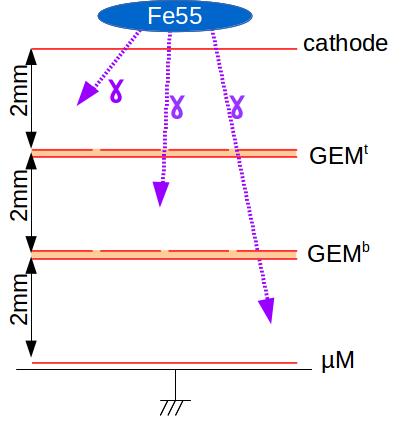}
    \includegraphics[width=0.55\linewidth]{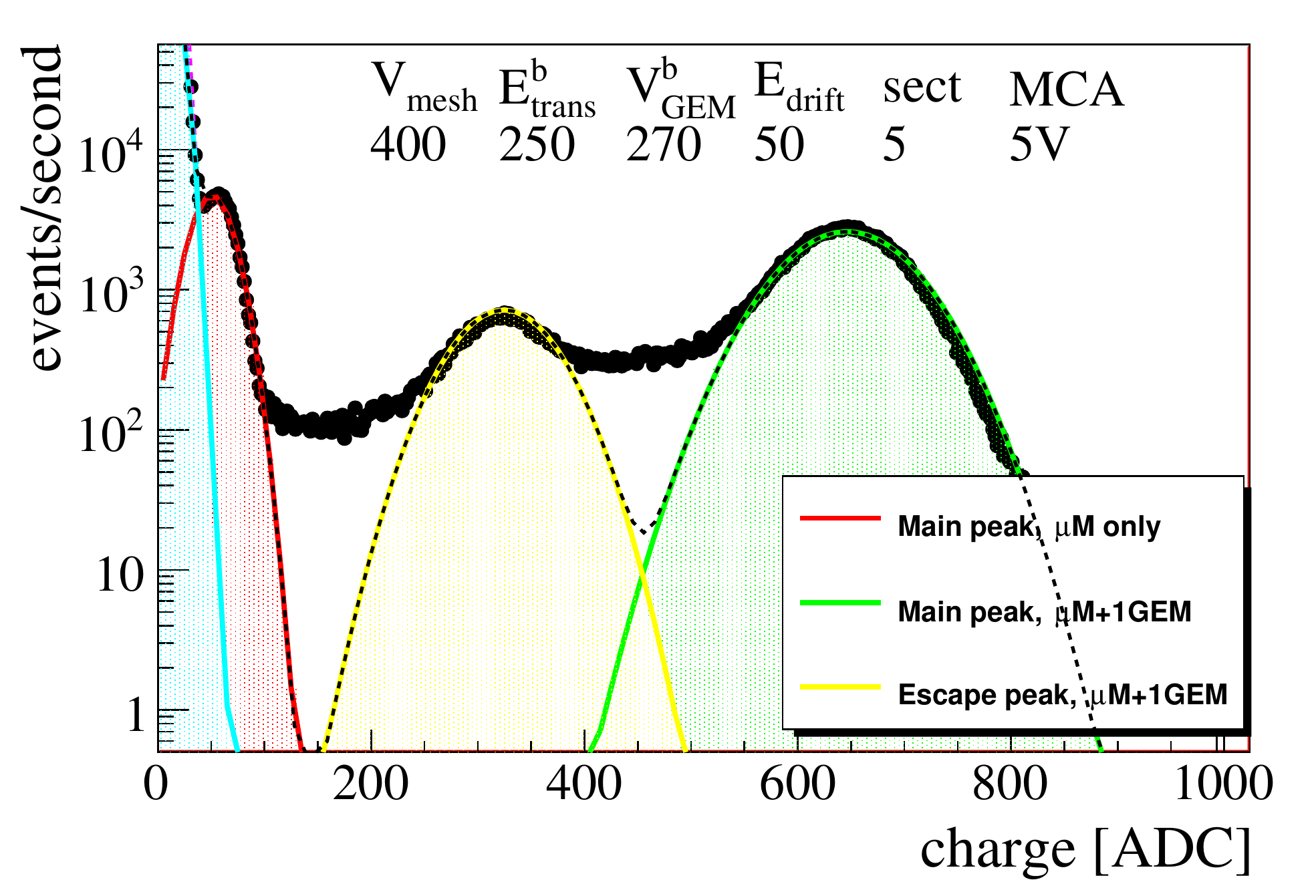}
    \caption{
       Left: the layout of the test setup, with one micromegas and two GEMs.
      (By applying a null or inverted voltage to the top GEM, we only observe conversion in the two lower regions).
      Right: an example of a measured $^{55}$Fe spectrum from amplification with one micromegas and one GEM.
      The main peak ($5.9\,\kilo\electronvolt$) is plotted after amplification with only the micromegas (red) or with both micromegas and GEM (green).
      The escape peak is only visible with full amplification (yellow).
    }
    \label{fig:TestBox}
  \end{center}
\end{figure}

We characterized the combination of a micromegas and either one or
two GEM in a gas mixture (Ar:95 \%Isobutane:5 \%) at atmospheric
pressure.
This was done in a dedicated test setup using a $^{55}$Fe
source\cite{PhGros:TIPP}.   
The successive amplification steps were kept at a distance of $2\,\milli\meter$ from each other by spacers (Fig.~\ref{fig:TestBox} left).

In argon, the X-rays from a $^{55}$Fe source deposit either
$5.9\,\kilo\electronvolt$ (main peak) or $2.7\,\kilo\electronvolt$
(escape peak) upon ionization.
This conversion can take place either above or below a given GEM sheet.
The ionization electrons are therefore either amplified by that GEM or
not.
A typical  spectrum is shown in Fig.~\ref{fig:TestBox} (right). 
The main and the escape peaks are seen with amplification from the
micromegas and one GEM, and the main peak with micromegas
amplification only.
The ratio of the position of the two main peaks provides a precise
measurement of the absolute GEM amplification gain.
Further details on these measurements, including foil transparency
and extraction efficiency, can be found in Ref.~\cite{PhGros:TIPP}.

\subsection{Characterization with cosmic rays}
\label{seb:sec:cosmics}

\begin{figure}[ht]
  \begin{center}
 \includegraphics[width=0.54\linewidth]{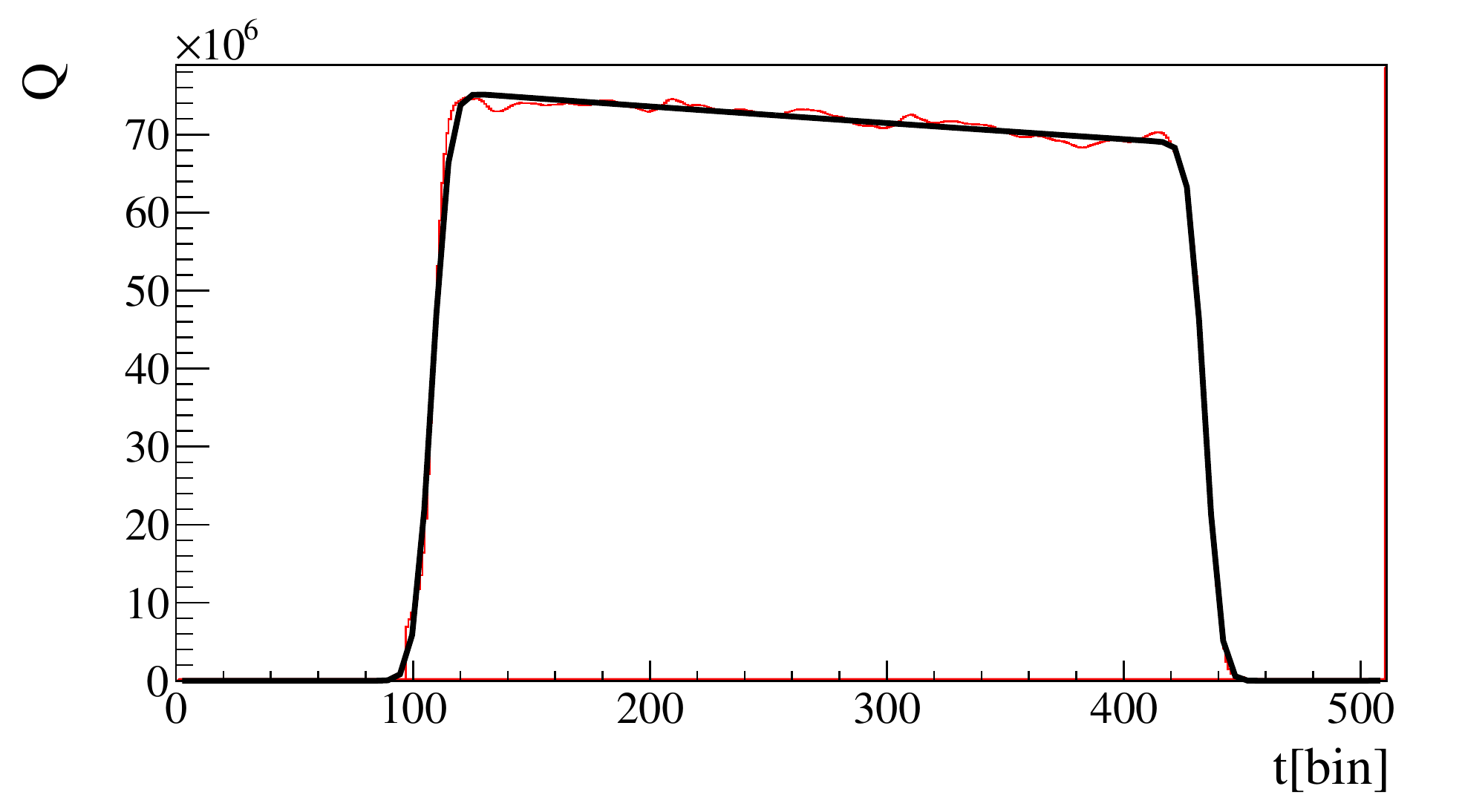}
 \includegraphics[width=0.44\linewidth]{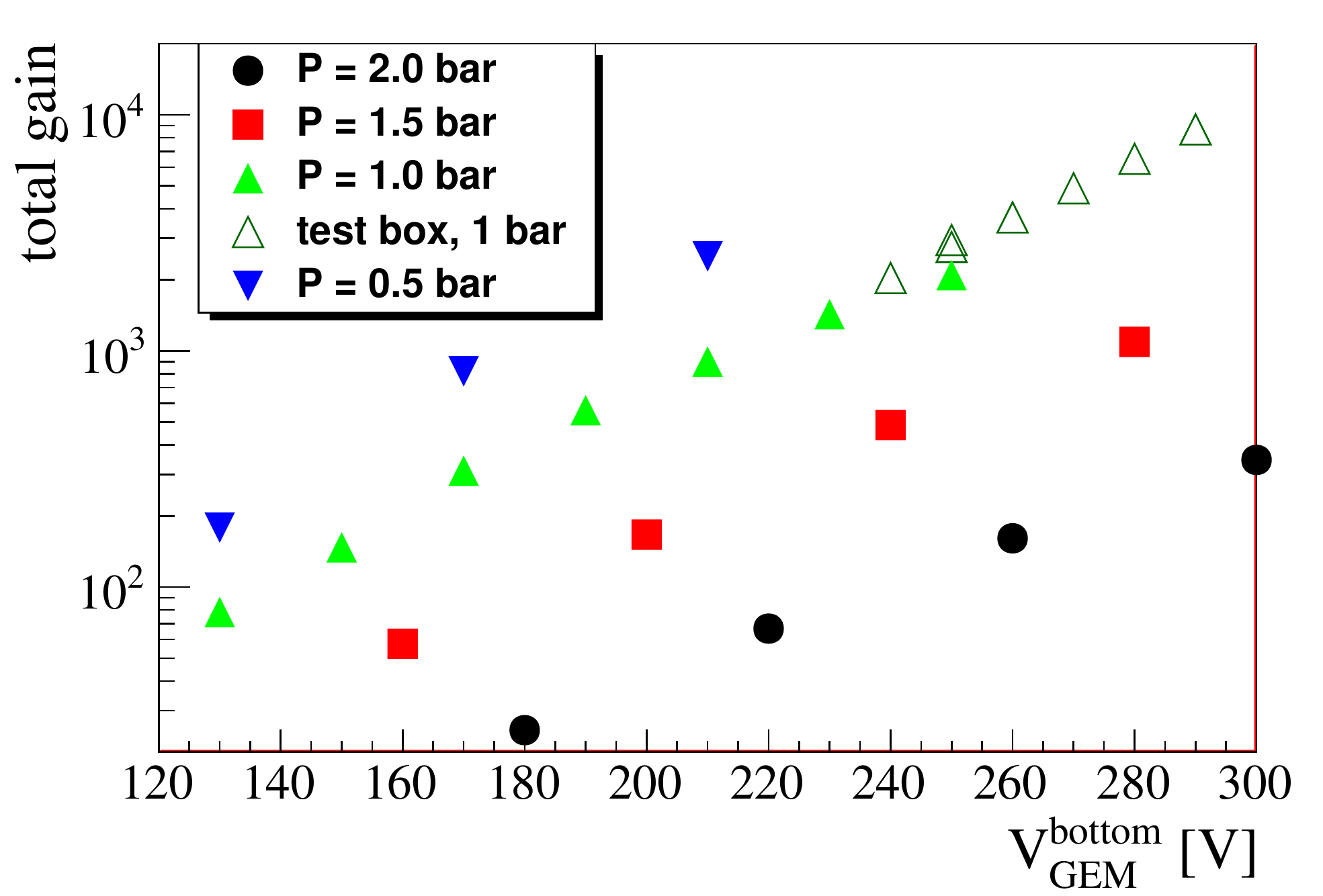}
    \caption{ Left: 
average total charge per track as a function of time, for a run of
cosmic rays that traversed the full TPC length, entering through the
anode and exiting through the anode.
We extract the  drift velocity,
the  electron absorption per unit length and
the amplification gain 
from such spectra.
Right: the effective gain measured from cosmic rays in the TPC for
several gas pressures.  The results at $1\,\bbar$ are consistent
with those obtained with the radioactive source.  }
    \label{fig:zDist}
  \end{center}
\end{figure}

The system of micromegas and 2 GEMs was commissioned in the
detector.
It was tested with cosmic rays, using the same gas mixture at several 
pressures.
Most cosmic muons are relativistic and therefore deposit the same average
energy per unit length.
The TPC was set "vertical", so that most cosmic rays entered it through
the amplification system.
We selected events with a cosmic ray that crossed the full $z$ of 
$30\, \centi\meter$ by triggering on the coincidence of the up and
down scintillator signals.

The charge distribution over drift time is shown in Fig.~\ref{fig:zDist} (left).
The value of the drift velocity, $v_{\textrm{drift}}$, was easily obtained using the basic relation of the TPC mechanism,
$z=v_{\textrm{drift}} \times t_{\textrm{drift}}$, since the cosmic rays traverse the full $z$ thickness of 
$L_{TPC}=30\,\centi\meter$.
The slope of the plateau corresponds to  electron absorption in the gas. 
The average charge per track for each run is used to estimate the amplification gain.
From this spectrum, the drift velocity, the attenuation length
and the total amplification gain were extracted.
Fig.~\ref{fig:zDist} (right) shows the amplification gain for several TPC
gas pressures as a function of the voltage on one of the
GEMs.
The measurements are consistent with those obtained using the
$^{55}$Fe source.

\section{Data-taking at NewSUBARU}

In Nov.~2014 the detector was exposed to a beam of pseudo-monochromatic gamma-ray
photons delivered by the BL01 beam line at NewSUBARU~\cite{NewSUBARU}.

\subsection{Laser-Compton Source (LCS)}

\begin{table}
\caption{\label{laser} Electron beam energy, laser wavelength, and $\gamma$-ray energy at the Compton edge obtained from these.}
\begin{center}
\begin{tabular}{cc|cccc|cc}
\br
\multicolumn{2}{c|}{$E_e^-$[$\giga\electronvolt$]}& 0.618 &  0.982 & 1.233 & 1.480 &\multicolumn{1}{c}{Pulsing rate}&\multicolumn{1}{c}{Polarization}\\\cline{1-6}
\multicolumn{1}{c}{Laser} & $\lambda$ [$\micro\meter$]&\multicolumn{1}{r}{}&$E_{\gamma}$&[$\mega\electronvolt$]&\multicolumn{1}{l|}{}&\multicolumn{1}{c}{[kHz]}&\\ 
\hline
Nd:YVO$_4$ ($2 \omega$) & 0.532 &  13.4 & 33.3  & 52.1 & 74.3  & 20 & $P \approx 1$ \\
Nd:YVO$_4$ ($1 \omega$) & 1.064 &  6.76 & 16.9  & 26.6 & 38.1  & 20 & $P \approx 1$ \\
Er(fiber)               & 1.540 & 4.68  & 11.8  & 18.5 & 26.5  & 200       & $P=0$  \\
CO$_2$                  & 10.55 &       & 1.74  & 2.73 & 3.93  & CW       & $P =0$ \\
\br
\end{tabular}
\end{center}
\end{table}

The gamma-ray beam  is produced by the inverse Compton
scattering of laser photons by relativistic electrons.
The electron beam energy can be varied in the range 
$0.6~\sim 1.5\,\giga\electronvolt$.
The laser beams available are: Nd:YVO$_4$ ($2 \omega$) with
wavelength $\lambda = 0.532\,\micro\meter$, Nd:YVO$_4$ ($1 \omega$)
$\lambda=1.064\,\micro\meter$, Er (fiber) $\lambda = 1.540\,\micro\meter$ and CO$_2$
$\lambda = 10.55\,\micro\meter$.
This results in a gamma energy range at the Compton edge between
$1.7\,\mega\electronvolt$ and $74\,\mega\electronvolt$ as shown in Table~\ref{laser}.
As the gamma has a maximum energy for forward Compton scattering, energy selection was performed by using a collimator on axis.
When collimation is applied, the polarisation of the laser beam is almost
entirely transferred to the gamma beam: in that way, an almost fully polarized gamma beam ($P \approx 1$) is obtained.
A general issue in polarization studies is the control of a possible
systematic bias induced by the non-cylindrical-symmetric structure
of the detector~\cite{Martin}.
Therefore in addition to the fully polarized data, some
data with a non polarized beam were taken.
To produce such a beam, a $\lambda/4$ plate is used. This changes the
linear polarization to circular polarization, which, as far as pair
conversion is concerned, is equivalent to random polarization ($P=0$).
Or vice versa.

\subsection{The HARPO Trigger system}

For the previous characterizations using cosmic rays, the
TPC was triggered on a simple coincidence of scintillator
signals~\cite{HARPO:Pisa2012,Bernard:2014kwa}.
In contrast, for data taking in beam, we wished to maximize the
fraction of selected gammas that converted in the gas. 
Therefore a dedicated, more sophisticated trigger was designed.
In addition it provides separately the control of the efficiency of
each of its components.

\subsubsection{Description of signals}

\begin{figure}[ht]
  \begin{center}
 \includegraphics[width=0.79\linewidth]{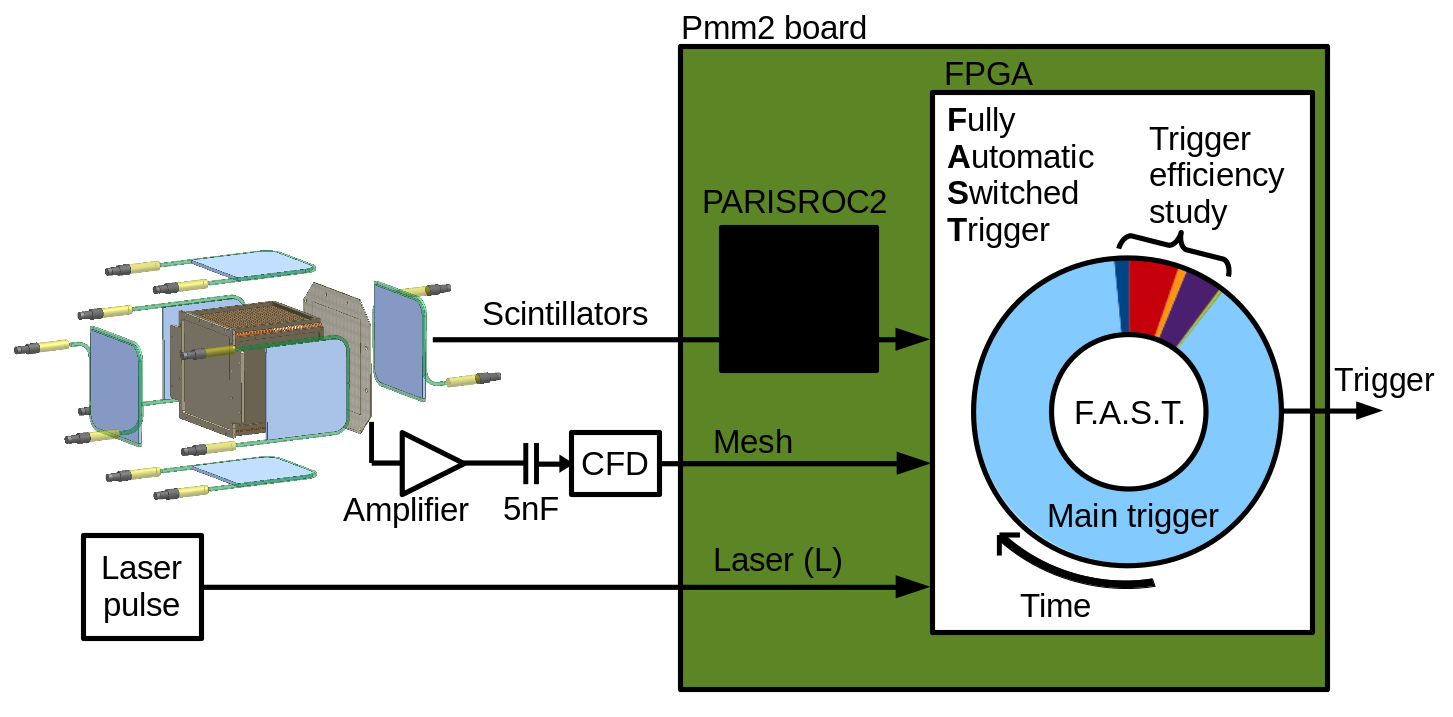}
    \caption{
      Global view of HARPO trigger system.
    }
    \label{fig:trigger}
  \end{center}
\end{figure}

The HARPO trigger system is based on a
PARISROC2~\cite{ConfortiDiLorenzo:2012sm} chip mounted on a
PMM2~\cite{Genolini:2008uc} board.
This trigger is built from the discriminated signals of the
scintillators, laser and micromegas mesh as shown in
Fig.~\ref{fig:trigger}.
The  available signals are: 

\begin{itemize}
\item Laser: when a pulsed laser is used, the laser's trigger
  signal $L$ (Fig.~\ref{fig:laser trigger}) is used.
It also determines the time at which the event takes place, $t_0$, with
a precision of a few $10$ ns.

\begin{figure}[ht]
  \begin{center}
 \includegraphics[width=0.68\linewidth]{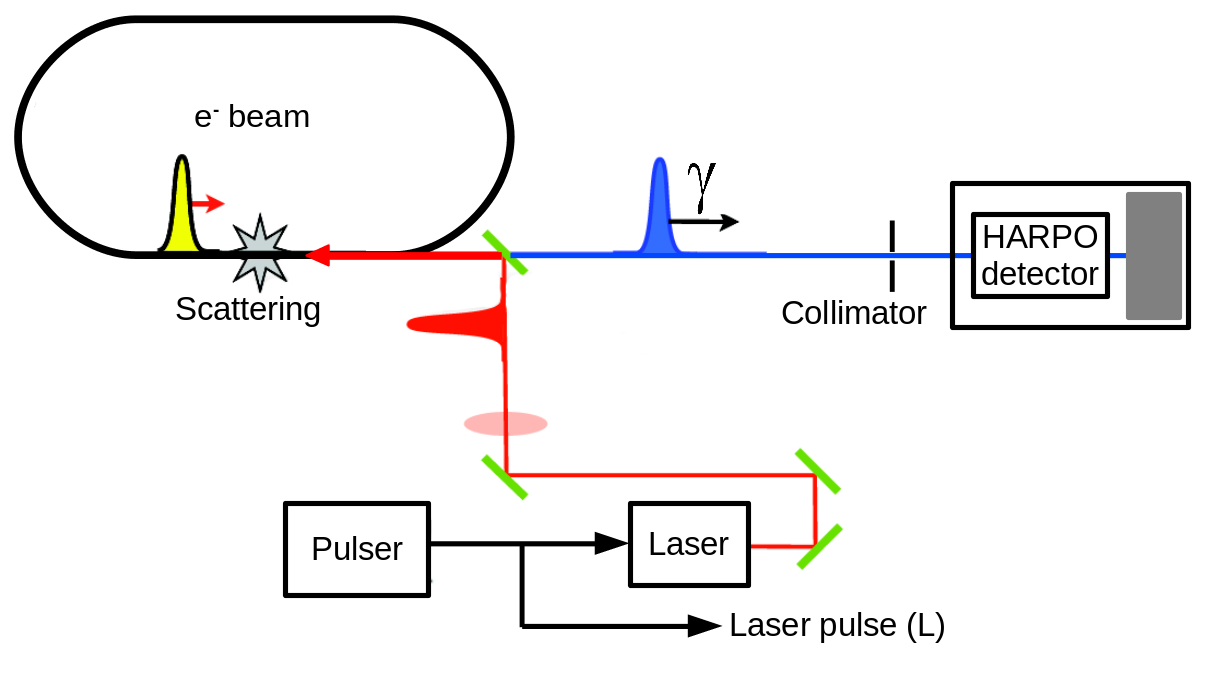}
    \caption{
      Schema of the Laser trigger used at NewSUBARU.
    }
    \label{fig:laser trigger}
  \end{center}
\end{figure}

\item Scintillator: the signal comes from 12 PMTs on 6 scintillators. 
The signal is recorded, whether there is a trigger or not.
It gives $t_0$ with a hundred nanosecond precision.
This signal determines $t_0$ when the laser signal is not available (CO$_2$).

\item Micromegas mesh: 
the signal induced on the mesh is long and has an unpredictable shape:
 it corresponds to the time distribution of the charge deposited by
the event in the TPC gas as it drifts towards the anode.  The signal is
amplified and derived through a 5 nF capacitor.
It is then discriminated with a constant-fraction discriminator (CFD), so as to
determine the rising edge  $t_{\textrm{start}}$ of the
signal.
The $RC$ constant of the readout electronics of the mesh signal is 
$\approx 1 \micro\second$.
It has been fixed as a compromise between the amount of electronic
noise given the large capacitance of the full mesh, of 8\,nF, and the
need of a precise determination of the rising edge of the signal  $t_{\textrm{start}}$.

Events that have a too small of a delay between the time of the event,
$t_0$, and the discriminated mesh signal rising edge,
$t_{\textrm{start}}$, are rejected.
This provides a veto on background tracks created in the beam line
upstream of the TPC and on gamma rays which converted in the material
between the up scintillator and the active TPC gas
 (example: in the PCB supporting the amplification system).

\end{itemize}

\subsubsection{Gamma trigger line.}

The gamma trigger, $T_\gamma$, is designed so as to select gamma rays produced by
LCS and that converted inside the TPC gas. The conditions are listed below:
\begin{itemize}
\item
the upstream scintillator signal ($S_{up}$) is used as a veto, 
\item
at least one downstream scintillator signal ($O$, other than ``up'') is required,
\item
the laser trigger signal ($L$), when present, is used,
\item
 the events with a signal present in the mesh are selected, vetoing the prompt part as already explained, we require $t_{\textrm{start}} - t_0 > 1 \micro\second$.
This ``slow'' selection of a mesh signal is denoted $M_{\textrm{slow}}$.

\end{itemize}

The main $\gamma$ trigger line is therefore defined as the combination
of the four following components:
\begin{equation}
T_\gamma \equiv 
\overline S_{up} \cap O \cap M_{\textrm{slow}} \cap L
\end{equation}

\begin{figure}[ht]
  \begin{center}
 \includegraphics[width=0.96\linewidth]{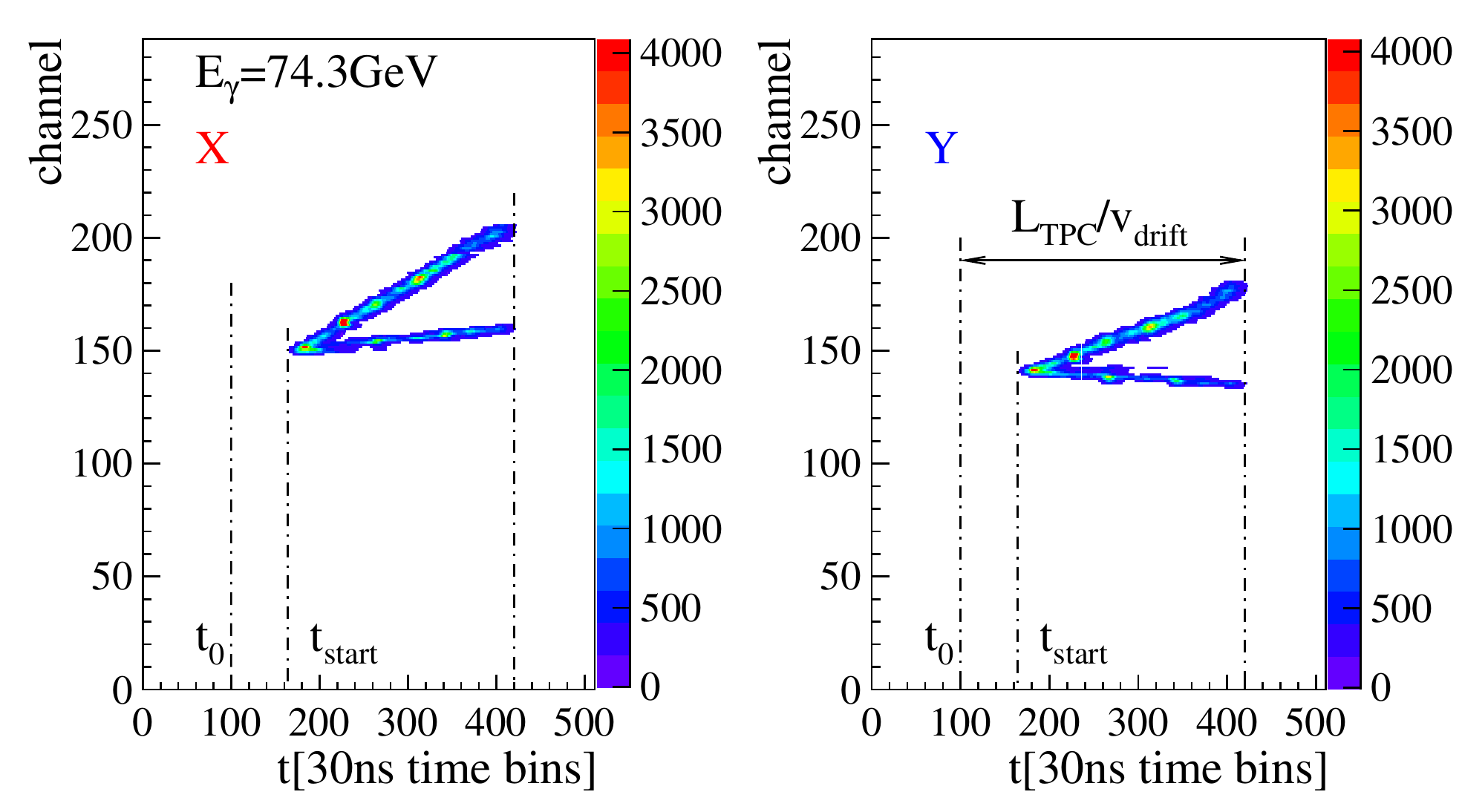}
    \caption{
      A gamma event that was selected by the gamma trigger, with  definitions of timing information.
    }
    \label{fig:gammatrigger}
  \end{center}
\end{figure}

\begin{figure}[ht]
  \begin{center}
 \includegraphics[width=0.9\linewidth]{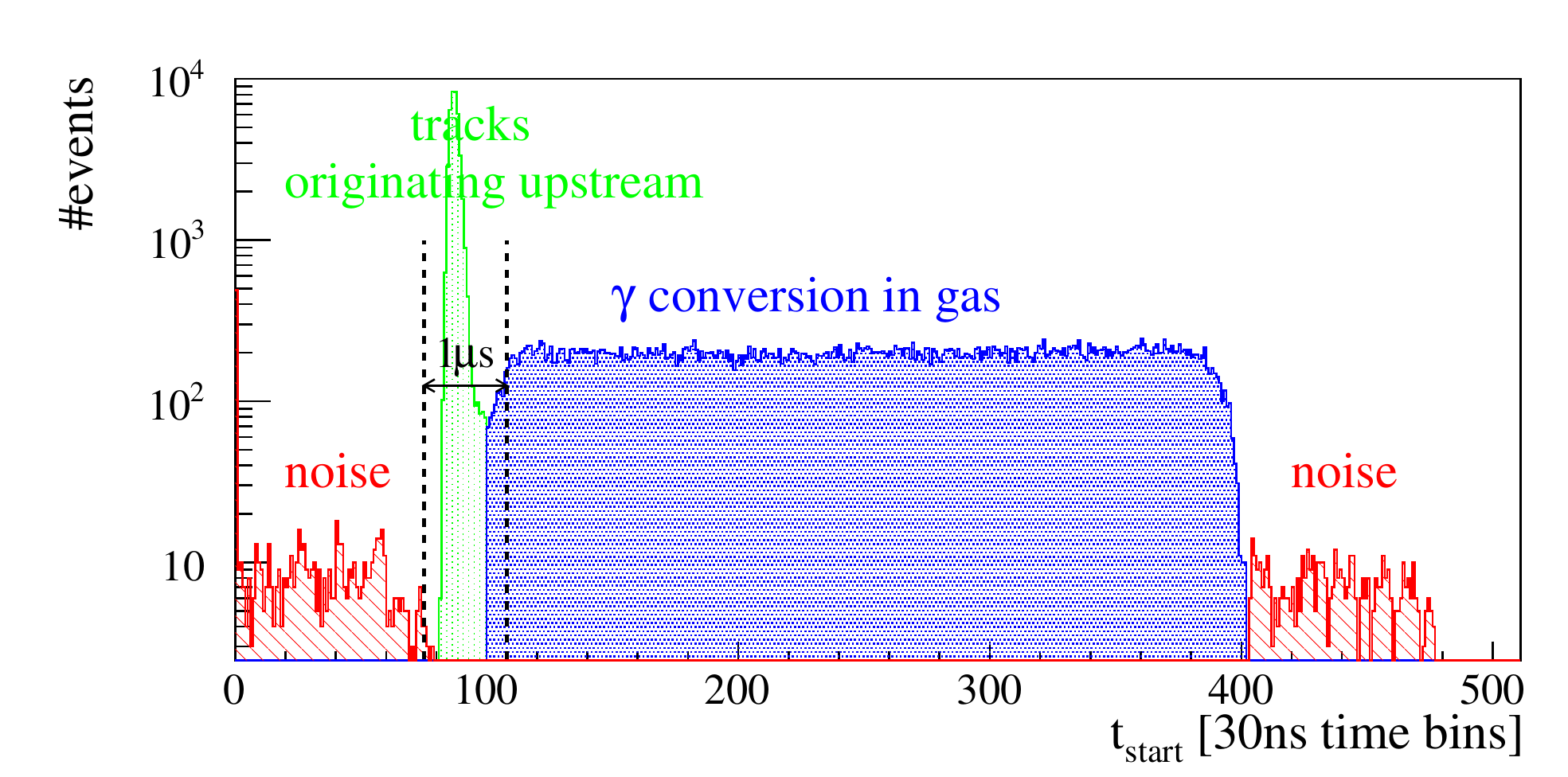}
    \caption{Distribution of 
      event rising edge $t_{\textrm{start}}$ for one run (all trigger lines): 
      tracks originating upstream (green), 
      gamma conversion inside the TPC gas (blue),
      charge deposition inside the gas by events not related to the trigger (red).
    }
    \label{fig:tmin}
  \end{center}
\end{figure}

The distribution of $t_{\textrm{start}}$ is shown in Fig.~\ref{fig:tmin}. 
Gamma conversions inside the TPC gas (blue) are the signal events. The
flatness of that part of the spectrum is due to the fact that the
probability of conversion per unit path length is  constant for a
thin detector.
Tracks entering the detector from upstream and gamma conversions in the
detector material upstream of the TPC gas, which escaped the vetos, form
the (green) peak. 
If we had used a trigger without any veto, the height of this peak
would have been larger by several orders of magnitude, precluding an efficient data taking.
Events (red) which lie outside the normal time range 
(i.e., $t_{\textrm{start}} <100$ or $t_{\textrm{start}} > 400$)
are due to energy deposition inside the gas by events that are not
related to the trigger (pile-up).

\subsubsection{Other Trigger Lines.}

In addition to the main gamma trigger line, separate trigger lines have been formed with one of the trigger components omitted
(such as 
$O \cap M_{\textrm{slow}} \cap L$, ~
$\overline S_{up} \cap M_{\textrm{slow}} \cap L$, ~
$\overline S_{up} \cap O \cap L$, ~
$\overline S_{up} \cap O \cap M_{\textrm{slow}}$,  ~
in the case of a pulsed laser).
Analysis of these specific data will enable us to control separately
the signal efficiency and the background rejection factor for each
component of the main trigger line.
Other dedicated lines were used, such as the ``traversing $z$ track'' line for calibration purposes.
Most of these additional lines were down-scaled so as to not saturate
the bandwidth of the digitizing electronics, and were active only
for a fraction of the time, as denoted by the rainbow part of Fig.
\ref{fig:trigger}.
Approximately $90\%$ of the time was devoted to the main trigger line.

\subsection{Gas}

The pressure vessel was evacuated down to $\sim 2 \times
10^{-6}\,\bbar$, rinsed with $\sim$ $0.1\,\bbar$ gas, evacuated
again, and then filled with $2.1\,\bbar$ of the gas mixture (Ar:95 \%
Isobutane:5 \%). 
This same gas was used in a sealed mode for 23 days.

At the end of the data taking, a pressure scan from 1 to 4 bar was performed, with
the gas amplification tuned so that the signal amplitude was kept
constant ($E_{e} = 1.5\,\giga\electronvolt$, Nd 1$\omega$, $P = 1$).

\subsection{Monitoring}

During data taking in the beam, some basic
parameters were monitored, such as alignment, amplification gain, trigger
performance and event quality after some very basic tracking.

\subsubsection{Alignment.}

\begin{figure} [Thb]
\begin{center}
 \includegraphics[width=0.96\linewidth]{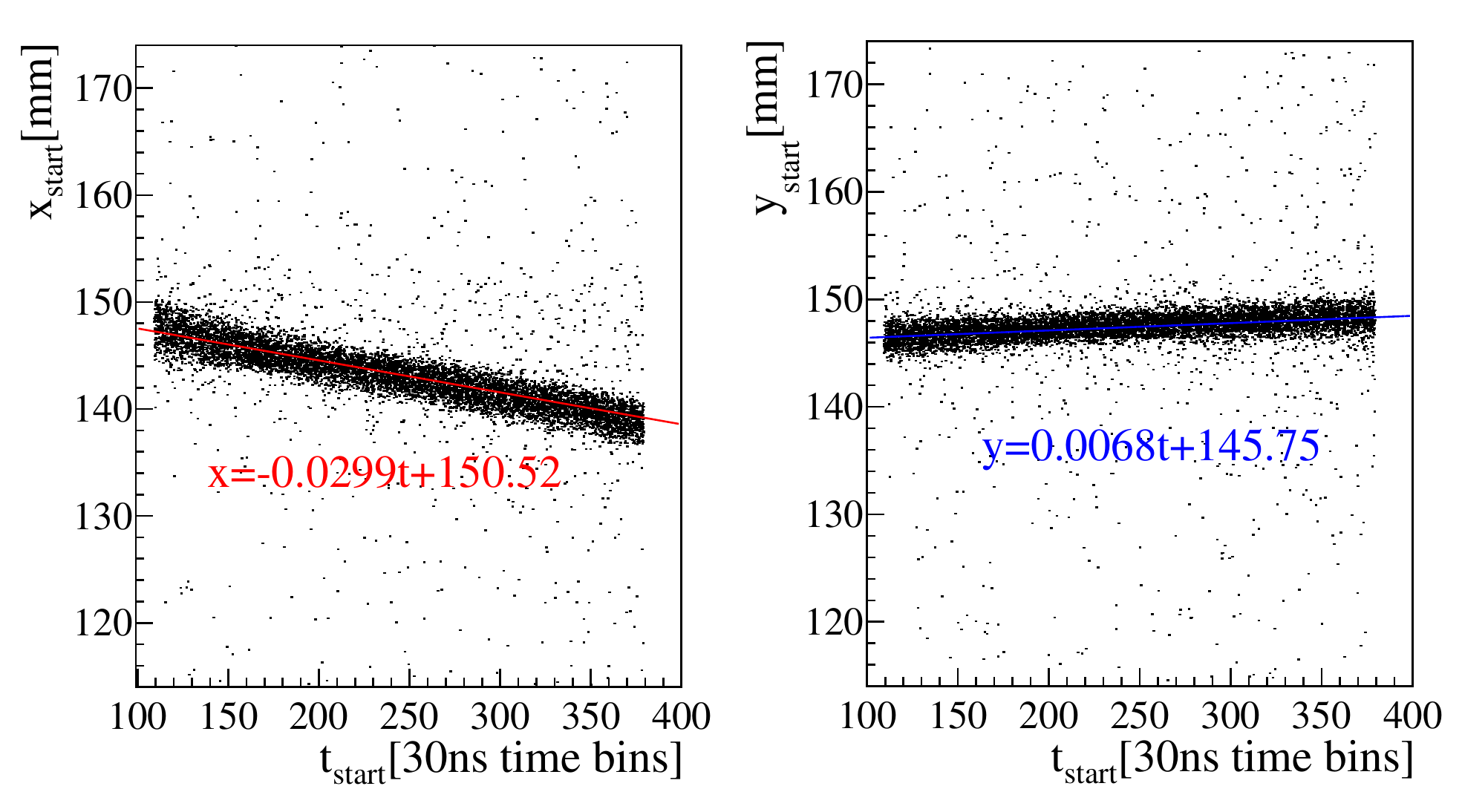}
\caption{Alignment plot for run 1277 (1\,mm diameter collimator), with linear fit
(gamma trigger line).
\label{fig:alignment:run1277lat}}
\end{center}
\end{figure}

The  alignment of the HARPO detector in the collimated gamma
beam was monitored by plotting the transverse ($x$ or $y$)
position of the gamma conversion candidate vertex as a function of
its longitudinal position $z$
(Fig. \ref{fig:alignment:run1277lat}).
The vertex is defined here from the charge-weighted barycenter of the
three first clusters of a selected gamma-conversion event (triggered by
the gamma trigger).

\subsubsection{Calibration.}

\begin{figure}[ht]
  \begin{center}
    \includegraphics[width=0.7\linewidth]{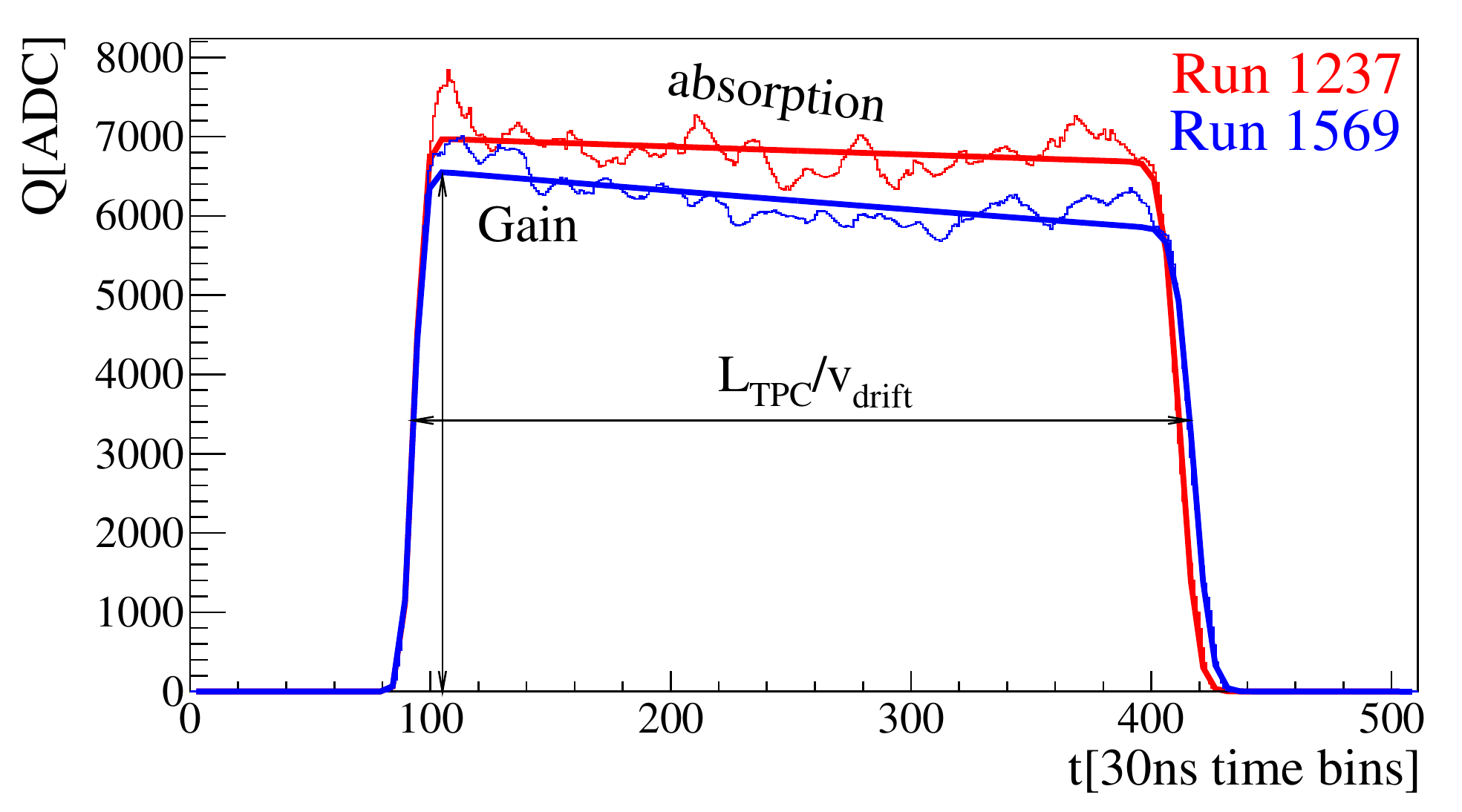}
    \caption{The average energy deposited at each time bin per
      traversing track along the $z$-axis for two runs separated by a
      full week of high flux running.  }
    \label{fig:charge-time}
  \end{center}
\end{figure}

The calibration of the TPC, i.e., the determination of the drift
velocity, of the amplification gain and of the
electron absorption, is performed by the analysis of the time
distribution of the average charge per track as was described in
section \ref{seb:sec:cosmics}.
A special trigger line ($\sim2 \%$ of the events) was used to
select the tracks traversing in the $z$ direction (i.e., along the
drift time) without biasing the time distribution.

\begin{figure}[ht]
  \begin{center}
 \includegraphics[width=0.98\linewidth]{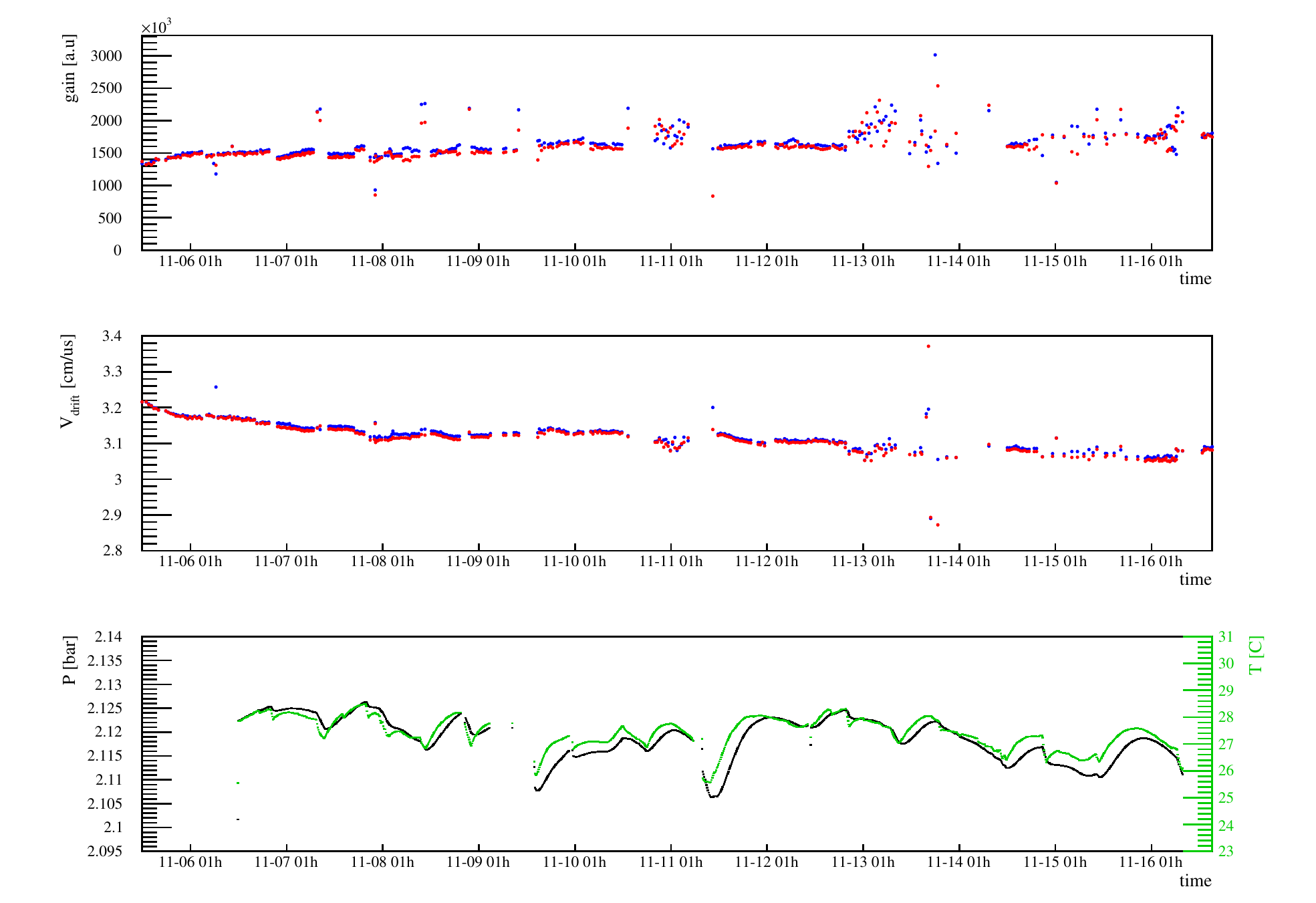}
    \caption{
 Time variation of the gain, estimated by the average charge per
 track, corrected for the track angle that is normalized to 30\,cm
 track length, the drift velocity ($x$ in red, $y$ in blue)
 and the pressure (black) and  temperature (green) of the gas.  }
    \label{fig:TimeEvolution}
  \end{center}
\end{figure}

Fig.~\ref{fig:charge-time} and Fig.~\ref{fig:TimeEvolution} show that
there was no significant gas deterioration over the two weeks in the beam.
The small features that can be seen in the time charts 
(Fig.~\ref{fig:TimeEvolution})
are associated
with changes in the gamma-beam energy and/or of the laser pulsing
rate, meaning that these small biases are related to the measurement,
not to the quantity measured.
The gas gain and the drift velocity were found to be sufficiently
stable and the electron absorption due to either leaks or outgassing
was found to be sufficiently small that we didn't have to renew the
gas.

\section{Conclusions and perspectives}

A gas TPC is the instrument suited to covering the performance gap in
cosmic gamma-ray detection in the MeV-GeV range.
We have built a demonstrator that includes a micromegas+GEM
amplification system.
We have performed an experimental campaign with this demonstrator
using the high-flux, quasi-monochromatic, polarised photon beam at
NewSUBARU.
Tests that were performed during data taking indicate an excellent
detector performance.

We have more than 1 TB of data ($>6 \times 10^{7}$ events, a large
fraction of which is estimated to be gamma conversions in the gas) to
analyze so that we can study the gamma conversion to e$^+$e$^-$ pairs and,
in particular, measure the low energy polarization asymmetry, and characterize
the performance of the demonstrator both as a gamma telescope and 
as a gamma polarimeter.
Further development is ongoing to meet the constraints of a space
environment~\cite{Bernard:2014kwa}, in particular, tests on the
behaviour of the detector in space with a dedicated trigger.
The next step will then be to verify its operation with a balloon
flight.

\section{Acknowledgments} 

This work is funded by the P2IO LabEx (ANR-10-LABX-0038) in the
framework ``Investissements d'Avenir'' (ANR-11-IDEX-0003-01) managed
by the French National Research Agency (ANR),
and directly by the ANR (ANR-13-BS05-0002).

This work was performed by using NewSUBARU-GACKO 
(Gamma Collaboration Hutch of Konan University).

\end{document}